\documentclass{llncs}
\usepackage{smwifclass}
\usepackage[utf8]{inputenc}
\usepackage{amsmath}
\usepackage{amsfonts}
\unlessclass{llncs}{\usepackage{amsthm}}
\usepackage{algorithm}
\usepackage{caption}
\usepackage[endLComment=~]{algpseudocodex}

\usepackage{balance}
\usepackage{xcolor}
\definecolor{darkblue}{rgb}{0.0,0.0,0.7}
\definecolor{darkgreen}{rgb}{0.0,0.5,0.0}
\usepackage{url}
\usepackage[colorlinks,bookmarks]{hyperref}
\hypersetup{linkcolor=darkblue,citecolor=darkblue,urlcolor=darkblue}
\usepackage{smwmath}
\usepackage{smwtools}
\showdetailstrue
\showanotatrue
\DeclareMathOperator{\lpquo}{lpquo}
\DeclareMathOperator{\lprem}{lprem}
\DeclareMathOperator{\rpquo}{rpquo}
\DeclareMathOperator{\rprem}{rprem}
\DeclareMathOperator{\lquo}{lquo}
\DeclareMathOperator{\lrem}{lrem}
\DeclareMathOperator{\rquo}{rquo}
\DeclareMathOperator{\rrem}{rrem}
\DeclareMathOperator{\iprec}{prec}
\DeclareMathOperator{\ishinv}{shinv}
\DeclareMathOperator{\ishift}{shift}
\DeclareMathOperator{\ilshinv}{lshinv}
\DeclareMathOperator{\ilshift}{lshift}
\DeclareMathOperator{\irshinv}{rshinv}
\DeclareMathOperator{\irshift}{rshift}
\newcommand{\NNInt}{\ensuremath{\mathbb Z_{\ge 0}}}

\newcommand{\scl}{\ensuremath{\Sc l}}
\newcommand{\scr}{\ensuremath{\Sc r}}

\newcommand{\iter}[1]{\ensuremath{_{(#1)}}}

\newcommand{\rcoeff}[2]{\ensuremath{#1_{#2}}}
\newcommand{\lcoeff}[2]{\ensuremath{\,{}_{#2}#1}}

\newcommand{\Rofx}{\ensuremath{R[x]}}
\newcommand{\Rskew}{\ensuremath{R[x; \sigma, \delta]}}

\newcommand{\Fofx}{\ensuremath{F[x]}}
\newcommand{\Sc}[1]{\ensuremath{\text{\sc #1}}}
\newcommand{\mtwo}[4]{\ensuremath{
\left [ \begin{array}{cc} #1 & #2 \\ #3 & #4 \end{array} \right ]
}}

\newcommand{\algotinyspace}{\upstrut{2.2ex}}
\newcommand{\algosmallspace}{\upstrut{2.8ex}}
\newcommand{\algomedspace}{\upstrut{3.5ex}}

\begin{document}
\title{Efficient Quotients of \mbox{Non-Commutative Polynomials}}
\author{Stephen M. Watt}
\newcommand{\theabstract}{
\begin{abstract}
It is shown how to compute quotients efficiently in non-commutative univariate polynomial rings.  
This extends earlier work where efficient generic quotients were studied with a primary focus on commutative domains.  
Fast algorithms are given for left and right quotients of polynomials where the variable commutes with coefficients.
These algorithms are based on the concept of the ``whole shifted inverse'', which is a specialized quotient where the dividend is a power of the polynomial variable.
It is also shown that when the variable does not commute with coefficients, that is for skew polynomials, left and right whole shifted inverses are defined and may be used to compute right and left quotients.  
In this case their computation is not asymptotically fast, but once obtained, they may be used to compute multiple quotients, each with one multiplication.
Examples are shown of polynomials with matrix coefficients, differential operators and difference operators.
In addition, a proof-of-concept generic Maple implementations is given.
\end{abstract}
}
\begin{IfClassLlncs}
  \institute{Cheriton School of Computer Science, University of Waterloo\\
  \url{https://cs.uwaterloo.ca/~smwatt} \\
  \email{smwatt@uwaterloo.ca}}
\end{IfClassLlncs}
\begin{IfClassArticle}
  \date{
    Cheriton School of Computer Science\\
    University of Waterloo\\
    \texttt{smwatt@uwaterloo.ca}
  }
\end{IfClassArticle}
\EarlyTitleAbstract
%%
%% Publisher Info %%%%%%%%%%%%%%%%%%%%%%%%%%%%%%%%%%%%%%%%%%%%%%%%%%%%%%%%%%%%%%%
%%
\LateTitleAbstract
\begin{IfClassLlncs}
\thispagestyle{plain}\pagestyle{plain}
\end{IfClassLlncs}
%23456789 123456789 123456789 123456789 123456789 123456789 123456789 123456789!

%%%%%%%%%%%%%%%%%%%%%%%%%%%%%%%%%%%%%%%%%%%%%%%%%%%%%%%%%%%%%%%%%%
\spacetune{\vspace{1cm}}
\section{Introduction}
\label{sec:introduction}
%%%%%%%%%%%%%%%%%%%%%%%%%%%%%%%%%%%%%%%%%%%%%%%%%%%%%%%%%%%%%%%%%%
In symbolic mathematical computation it is important to have efficient algorithms for the fundamental arithmetic operations of addition, multiplication and division.   While linear time algorithms for additive operations are usually straightforward, considerable attention has been devoted to find efficient methods to compute products and quotients of integers, polynomials with integer or finite field coefficients and matrices with elements from a ring.  For these, both practically efficient algorithms and theoretically important bounds are well known.

For integer and polynomial division, efficient algorithms based on Newton iteration allow the computation of quotients in time proportional to multiplication.
Until recently, these algorithms left the original domain to perform arithmetic in related domains.
For integers, this involved computing an approximation to the inverse of the divisor in extended precision approximate arithmetic or in a residue ring, and for polynomials it involved computing the inverse of the reverse of the divisor polynomial in ideal-adic arithmetic.

We have recently shown how these quotients may be computed without leaving the original domain, and we have extended this to a generic domain-preserving algorithm for rings with a suitable whole shift operation~\cite{watt-issac-2023}.
For integers the whole shift multiplies by a power of the representation base and for polynomials it multiplies by a power of the variable, in both cases discarding terms with negative powers.
The previous paper developed the concept of the whole shifted inverse and used it to compute quotients efficiently.  Non-commutative domains were mentioned only briefly.

The present article expands on how these methods may be used to compute quotients of non-commutative polynomials.  
In particular, it is shown that
\begin{itemize}
    \item[$\bullet$] the whole shifted inverse is well-defined on non-commutative polynomial rings $R[x]$, 
    \item[$\bullet$] its computation is efficient,
    \item[$\bullet$] they may be used to compute left or right quotients in $R[x]$, each with one multiplication,
    \item[$\bullet$] left and right whole shifted inverses may be defined on skew polynomials $R[x; \sigma, \delta]$, and
    \item[$\bullet$] they may be used to compute the right and left quotients in $R[x; \sigma, \delta]$, each with one multiplication.
\end{itemize}

The remainder of this article is organized as follows.
Section~\ref{sec:background} presents some basic background, including notation, the definition of division in a non-commutative context,
and the Newton-Schulz iteration.
Section~\ref{sec:r-poly-division} considers division of non-commutative polynomials in $R[x]$, showing $O(n^2)$ algorithms for classical division and for pseudodivision. 
It recalls the notion of the whole shifted inverse, proves it is well-defined on non-commutative $R[x]$ and shows that it can be used to compute left and right quotients in this setting.
Section~\ref{sec:generic-algo} recapitulates the generic algorithms from~\cite{watt-issac-2023} that use a modified Newton iteration to compute the whole shifted inverse. 
It also explains why it applies when polynomial coefficients are non-commutative.
Section~\ref{sec:r-poly-example} gives an example of these algorithms applied to polynomial matrices.
Section~\ref{sec:r-skew-division} extends the discussion to skew polynomials $R[x;\sigma, \delta]$, defining left and right whole shifted inverse, and showing how they may be used. 
Section~\ref{sec:r-skew-example} gives linear ordinary differential and difference operators as examples, before concluding remarks in Section~\ref{sec:conclusions}.

%%%%%%%%%%%%%%%%%%%%%%%%%%%%%%%%%%%%%%%%%%%%%%%%%%%%%%%%%%%%%%%%%%
\spacetune{\vfill\pagebreak}
\section{Background}
\label{sec:background}
%%%%%%%%%%%%%%%%%%%%%%%%%%%%%%%%%%%%%%%%%%%%%%%%%%%%%%%%%%%%%%%%%%
\newcommand{\smallx}{{\bf \footnotesize x}}
\newcommand{\tablesep}{\upstrut{3ex}}
\subsection{Notation}
We adopt the following notation:\\
\noindent
\begin{tabular}{@{}cl@{}}
$\iprec_B u$
    & number of base-$B$ digits of an integer $u$, $\floor{\log_B |u|} + 1$ \tablesep\\
$\iprec_x p$
    & number of coefficients of a polynomial $p$, $\pdegree_x p + 1$
    \tablesep\\
$ u \iquo v,\; u \irem v$ 
    & quotient and remainder (see below) 
    \tablesep\\
$ u\, \text{\smallx quo}\, v,\; u\, \text{\smallx rem}\, v$ 
    & left and right (pseudo)quotient and remainder, $\text{\smallx} \in \{\mathrm{l, lp, r, pr} \}$
    \tablesep\\
$\ishift_n v$, $\ishinv_n v$
    & whole shift  and whole shifted inverse (see below)
    \tablesep\\
$R[x; \sigma, \delta]$, $R[x, \delta]$
    & skew polynomials (see Section~\ref{sec:r-skew-division}) 
    \tablesep\\
$\lcoeff ui$, $\rcoeff ui$
    & coefficient of skew polynomial $u$ with variable powers on the left, right.
    \tablesep\\
$\text{\smallx shift}_n v$, $\text{\smallx shinv}_n v$
    & left and right whole shift and shifted inverse, $\text{\smallx} \in \{\mathrm{l, r} \}$ (see Section~\ref{sec:r-skew-division})
    \tablesep\\
$X\iter i$  
    & value of $X$ at $i^{th}$ iteration
    \tablesep
\end{tabular}
\\[\baselineskip]
\noindent
The ``$\iprec$'' notation, standing for ``precision'', means the number of base-$B$ digits or polynomial coefficients. It is similar to that of~\cite{Moenck1972}, where it is used to present certain algorithms generically for integers and polynomials.
In particular, if we take integers to be represented in base-$B$,  \textit{i.e.} for any integer $u \ne 0$ there is $h = \iprec_B(u)-1$, such that
\begin{equation}
u = \sum_{i=0}^h u_i B^i, \quad u_i \in \mathbb Z,\, 0 \le u_i < B, \; u_h \ne 0,
\label{eq:base-B}
\end{equation}
then integers base-$B$ behave similarly to univariate polynomials with coefficients $u_i$, but with carries complicating matters.

\subsection{Division}
\label{sec:division-background}
The notion of integer quotients and remainders can be extended to more general rings.
For a Euclidean domain $D$ with valuation $N: D \rightarrow \NNInt$, such that for any $u, v \in D, v \ne 0$, there exist $q, r \in D$ such that 
\begin{align*}
u &= q v + r, & r &= 0 \text{ or } N(r) < N(v).
\end{align*}
The value $q$ is a \textit{quotient} of $u$ and $v$ and $r$ is a \textit{remainder} of dividing $u$ by $v$ and we write 
\begin{align*}
q &= u \iquo v & r &= u \irem v
\end{align*}
when these are unique.   When both the quotient and remainder are required, we write $u~\mathrm{div}~v = (u~ \iquo~ v, u~ \irem~ v)$.
When $D$ is a non-commutative ring with a valuation $N$, there \textit{may} exist 
left and right quotients such that
\begin{equation}
\begin{aligned}
    u &= v \,q_\scl + r_\scl, &\hspace{1.5cm} r_\scl &= 0 \text{ or } N(r_\scl) < N(v) \\
    u &= q_\scr \,v + r_\scr, &\hspace{1.5cm} r_\scr &= 0 \text{ or } N(r_\scr) < N(v).
\end{aligned}
\label{eq:lrquo-def}
\end{equation}
When these exist and are unique, we write
\begin{align*}
    q_\scl &= u \lquo v & r_\scl &= u \lrem v & % \\
    q_\scr &= u \rquo v & r_\scr &= u \rrem v .
\end{align*}

For certain non-commutative rings with a distance measure $\| \cdot \|$, a sequence of approximations to the inverse of $A$ may be computed via the Newton-Schulz iteration~\cite{schulz-1933}
\begin{equation}
X\iter{i+1} = X\iter i + X\iter i (1 - A X\iter i)
\label{eq:newton-schulz-iteration}
\end{equation}
where $1$ denotes the multiplicative identity of the ring.
There are several ways to arrange this expression, but the form above 
emphasizes that as $X\iter i$  approaches $A^{-1}$, the product $X\iter i(1 - AX\iter i)$ approaches $0$.
For $\mathbb C^{n\times n}$ matrices, a suitable initial value is
\(X\iter 0 = A^\dagger/(n\, \mathrm{Tr}(A A^\dagger))\),
where $A^\dagger$ is the Hermitian transpose.

\subsection{Whole Shift and Whole Shifted Inverse}
\label{sec:whole-shifted-inverse}
In previous work~\cite{watt-issac-2023} we studied the problem of efficient domain-preserving computation of quotients and remainders for integers and polynomials, then generalized these results to a generic setting.   To this end, we defined the notions of the \textit{whole shift} and \textit{whole shifted inverse} with attention to commutative domains.  We recapitulate these definitions and two results relevant to the present article.

\begin{Definition}[Whole $n$-shift in \Rofx]
Given a polynomial $u = \sum_{i=0}^h u_ix^i \in R[x]$, with $R$ a ring and $n \in \mathbb Z$,  the \emph{whole $n$-shift of $u$ with respect to $x$} is
\begin{equation}
\ishift_{n,x} u = \sum_{i+n\ge 0} u_i x^{i+n}.
    \label{defn:whole-shift-Rx}
\end{equation}
When $x$ is clear by context, we write $\ishift_n u$.
\end{Definition}
\begin{Definition}[Whole $n$-shifted inverse in $\Fofx$]
Given $n \in \NNInt$ and $v\in \Fofx$, $F$ a field, the \emph{whole  $n$-shifted inverse of $v$ with respect to $x$} is
\begin{equation}
   \ishinv_{n,x} v = x^n \iquo v.
    \label{defn:whole-shinv-Rx}
\end{equation}
When $x$ is clear by context, we write $\ishinv_n v$,
\end{Definition}
\begin{Theorem}
Given two polynomials $u, v \in \Fofx$, $F$ a field, and $0 \le \pdegree u \le h$,
\begin{equation}
u \iquo v = \ishift_{-h}(u \cdot \ishinv_h v).
\label{eq:quo-by-shinv}
\end{equation}
\end{Theorem}
For classical and Karatsuba multiplication 
it is more efficient to compute just the top part of the product in~\eqref{eq:quo-by-shinv}, omitting the lower $h$ terms, instead of shifting:
\[
\ishift_{-h}(u \cdot \ishinv_h v) = \Sc{MultQuo}(u, \ishinv_h v, h),
\]
with $\Sc{MultQuo}(a,b,n) = ab \iquo x^n$ computing only $\pdegree a + \pdegree b - n + 1$ terms.
For multiplication methods where computing only the top part of the product gives no saving, some improvement is obtained using
\[\ishift_{-h}(u \cdot \ishinv_h v) = \ishift_{-(h-k)}(\ishift_{-k} u \cdot \ishinv_h v). \]
\begin{Theorem}
Given $v \in F[x]$, with $F$ a field and $h > \pdegree v = k$ and suitable starting value $w\iter 0$, the sequence of iterates
\[
w\iter{i+1} = w\iter i + \ishift_{-h} \big ( w\iter i (\ishift_h 1 - vw\iter i ) \big )
\]
converges to $\ishinv_h v$ in $\lceil \log_2(h-k) \rceil$ steps.
\end{Theorem}
A suitable starting value for $w\iter 0$ is given by $\Sc{Shinv0}$ in Section~\ref{sec:generic-algo}.

\begin{algorithm}[t]
\caption{Classical division for non-commutative $R[x]$ with invertible $v_k$\algotinyspace}
\label{algo:classical-non-commutative-polynomial-quotient}
\begin{algorithmic}[1]
\LComment{Compute $q = \sum_{i=0}^{h-k}q_i x^i$ and $r = \sum_{i=0}^{k-1}r_i x^i$ such that $u = q \times_\pi v + r.$\algomedspace}
\Function{\Sc{div}\,}{$u = \sum_{i=0}^h u_i x^i\in R[x], v  = \sum_{i = 0}^k v_i x^i \in R[x], \pi\in S_2$}\algomedspace
    \State $v^* \gets \mathrm{inv}~v_k$
    \State $q \gets 0$
    \State $r \gets u$
    \For{$i \gets h-k$ to $0$ by $-1$}
        \State $t \gets (r_{i+k} \times_\pi v^*)\, x^i$
        \State $q \gets q + t$ 
        \State $r \gets r - t \times_\pi v$
    \EndFor
    \State \Return (q, r)
\EndFunction
\LComment{Left division:~\, 
$(q_{\Sc l}, r_{\Sc l}) \gets \Sc{ldiv}(u,v) 
\Rightarrow  u = v \times q_{\Sc l} + r_{\Sc l}$
\algomedspace}
\State $\Sc{ldiv}(u, v) \mapsto \Sc{div}\big ( u, v, (2\, 1) \big )$
\LComment{Right division: 
$(q_{\Sc r}, r_{\Sc r}) \gets \Sc{rdiv}(u,v) 
\Rightarrow  u = q_{\Sc r} \times v + r_{\Sc r}$
\algomedspace}
\State $\Sc{rdiv}(u, v) \mapsto \Sc{div}\big ( u, v, (1\, 2) \big )$
\end{algorithmic}
\end{algorithm}
%%%%%%%%%%%%%%%%%%%%%%%%%%%%%%%%%%%%%%%%%%%%%%%%%%%%%%%%%%%%%%%%%%
\section{Division in Non-Commutative $R[x]$}
\label{sec:r-poly-division}
%%%%%%%%%%%%%%%%%%%%%%%%%%%%%%%%%%%%%%%%%%%%%%%%%%%%%%%%%%%%%%%%%%
We now lay out how to use $\ishift$ and $\ishinv$ to compute quotients for polynomials with non-commutative coefficients.
First we  show classical algorithms to compute left and right quotients in $R[x]$.
We then prove two theorems, one showing that $x^n \lquo v = x^n \rquo v$ in this setting, making the whole shifted inverse well defined, and another showing that it may be used to compute left and right quotients.

\subsection{Definitions and Classical Algorithms}
Let $u$ and $v$ be two polynomials in $R[x]$ with Euclidean norm being the polynomial degree.
The left and right quotients and remainders are defined as in~\eqref{eq:lrquo-def}.
Left and right quotients will exist provided that $v_k$ is invertible in $R$ and they may be computed by Algorithm~\ref{algo:classical-non-commutative-polynomial-quotient}.  
In the presentation of the algorithm, $\pi$ denotes a permutation on two elements so is either the identity or a transposition.  
The notation $\times_\pi$ is a shorthand for $\times \circ \pi$ so $a \times_\pi b = a \times b$ when $\pi$ is the identity and $a \times_\pi b = b \times a$ when $\pi$ is a transposition.

There are some circumstances where quotients or related quantities may be computed even if $v_k$ is not invertible.
When $R$ is an integral domain, quotients may be computed as usual in $K[x]$ with $K$ being the quotient field of $R$.  
Alternatively, when $R$ is non-commutative but $v_k$ commutes with $v$, it is possible to compute \textit{pseudoquotients} and \textit{pseudoremainders} satisfying
\begin{align*}
m \, u &= v \, q_\scl + r_\scl,  &\pdegree r_\scl &< \pdegree v \\
u \, m &= q_\scr \, v + r_\scr,  &\pdegree r_\scr &< \pdegree v \\
m &= v_k^{h-k+1},
\end{align*}
as shown in Algorithm~\ref{algo:classical-non-commutative-polynomial-pseudoquotient}.
In this case, we write
\begin{align*}
    q_\scl &= u \lpquo v & r_\scl &= \lprem v \\
    q_\scr &= u \rpquo v & r_\scl &= \rprem v.
\end{align*}
Requiring $v_k$ to commute with $v$ is quite restrictive, however, so we focus our attention to situations where the inverse of $v_k$ exists.
\begin{algorithm}[t]
\caption{Non-commutative polynomial pseudodivision\algotinyspace}
\label{algo:classical-non-commutative-polynomial-pseudoquotient}
\begin{algorithmic}[1]
\LComment{Compute $q = \sum_{i=0}^{h-k}q_i x^i$ and $r = \sum_{i=0}^{k-1}r_i x^i$ such that $v_k^{h-k+1} u = q \times_\pi v + r$.\algomedspace\newline
Requires $v \times v_k = v_k \times v$.
}
\Function{\Sc{pdiv}\,}{$u = \sum_{i=0}^h u_i x^i\in R[x], v  = \sum_{i = 0}^k v_i x^i \in R[x], \pi\in S_2$}\algomedspace
    \State $q \gets 0$
    \State $r \gets u$
    \For{$i \gets h-k$ to $0$ by $-1$}
        \State $t \gets u_{i+k} \, x^i$
        \State $q \gets q + t \times_\pi v_k^i$
        \State $r \gets r\times_\pi v_k- t \times_\pi v$
    \EndFor
    \State \Return (q, r)
\EndFunction
\LComment{Left pseudodivision:~\, 
$(q_{\Sc l}, r_{\Sc l}) \gets \Sc{lpdiv}(u,v) 
\Rightarrow  v_k^{h-k+1} u = v \times q_{\Sc l} + r_{\Sc l}$
\algomedspace}
\State $\Sc{lpdiv}(u, v) \mapsto \Sc{pdiv}\big ( u, v, (2\, 1) \big )$
\LComment{Right pseudodivision: 
$(q_{\Sc r}, r_{\Sc r}) \gets \Sc{rpdiv}(u,v) 
\Rightarrow  v_k^{h-k+1} u = q_{\Sc r} \times v + r_{\Sc r}$
\algomedspace}
\State $\Sc{rpdiv}(u, v) \mapsto \Sc{pdiv}\big ( u, v, (1\, 2) \big )$
\end{algorithmic}
\end{algorithm}

\subsection{Whole Shift and Whole Shifted Inverse in $\Rofx$}
We now examine the notions of the whole shift and whole shifted inverse for $R[x]$ with non-commutative $R$.  First consider the whole shift. Since $x$ commutes with all values in $R[x]$, we may without ambiguity take, for $u = \sum_{i = 0}^h u_i x^i$ and $n \in \mathbb Z$,
\begin{equation}
\ishift_n u \;= \sum_{i +n \ge 0} x^n (u_i x^i)
            \;= \sum_{i +n \ge 0} (u_i x^i) x^n.
\end{equation}
That is, the fact that $R[x]$ is non-commutative does not lead to left and right variants of the whole shift.

\spacetune{\pagebreak}
\noindent
We state two simple theorems with obvious proofs:
\begin{Theorem}
Let $w \in R[x]$. Then, for all $n \in \NNInt$, 
\(
\ishift_{-n} \ishift_n w = w.
\)
\label{thm:shift-cancel}
\end{Theorem}
\begin{Theorem}
Let $u, v \in R[x]$ with $\pdegree u = h$ and $\pdegree v = k$.
Then, for $m \in \mathbb Z$,
\begin{align*}
\ishift_{-k-m} (u \times v)   &= \ishift_{-k} (\ishift_{-m}(u) \times v) \\
\ishift_{-h-m} (u \times v)   &= \ishift_{-h} (u \times \ishift_{-m}(v)).
\end{align*}
\label{thm:shift-factor}
\end{Theorem}%

We now come to the main point of this section and show $\ishinv$ is well-defined when $R$ is non-commutative.
\begin{Theorem}[Whole shifted inverse for non-commutative $\Rofx$]~\newline
Let $v = \sum_{i=0}^k v_i x^i \in R[x]$, with $R$ a non-commutative ring and $v_k$ invertible in $R$.   Then, for $h \in \NNInt$,
\[
x^h \lquo v  = x^h \rquo v.
\]
\label{thm:r-poly-shinv-ok}
\end{Theorem}
\begin{Proof}
Let $q_\scl = x^h \lquo v$ and $q_\scr = x^h \rquo v$.  
% We show that this implies $q_\scl = q_\scr$.
If $h < k$, then $q_\scl = q_\scr = 0$.
Otherwise, both $q_\scl$ and $q_\scr$ have degree $h-k \ge 0$
so
\begin{align}
v_k \, {q_\scl}_{h-k} &= 1 &
{q_\scr}_{h-k} \, v_k &= 1
\label{eq:qhmk}
\\
\sum_{j=M}^k v_j \,{q_\scl}_{i+k-j} &= 0 &
\sum_{j=M}^k {q_\scr}_{i+k-j} \,v_j &= 0,
\quad 0 \le i < h-k,
\label{eq:qsum}
\end{align}
where $M=\max(0, i -h +2k)$.   
We show by induction on $i$ that ${q_\scl}_i = {q_\scr}_i$ for $0 \le i \le h-k$.
Since $v_k$ is invertible, \eqref{eq:qhmk} and \eqref{eq:qsum} give
\begin{equation}
{q_\scl}_{h-k} =  {q_\scr}_{h-k} = v_k^{-1}
\label{eq:shinv-induction-base}
\end{equation}
and 
\begin{align}
{q_\scl}_i &= -\sum_{j=M}^{k-1} v_k^{-1}\, v_j \, {q_\scl}_{i+k-j}
&
{q_\scr}_i &= -\sum_{j=M}^{k-1} {q_\scr}_{i+k-j} \,v_j \, v_k^{-1} 
, \quad 0 \le i < h-k.
\label{eq:qsum-massage}
\end{align}
Equation~\eqref{eq:shinv-induction-base} gives the base of the induction.
Now suppose ${q_\scl}_i = {q_\scr}_i$ for $N < i \le h-k$.  Then for $i = N \ge 0$ equation \eqref{eq:qsum-massage} gives
\begin{align*}
{q_\scl}_N &= -\sum_{j=M}^{k-1} v_k^{-1}\,v_j \, {q_\scl}_{N+k-j}
            = -\sum_{j=M}^{k-1} v_k^{-1}\,v_j \, {q_\scr}_{N+k-j} \\
           &= -\sum_{j=M}^{k-1}  
v_k^{-1}\,v_j \left ( -\sum_{\ell=M}^{k-1} \, {q_\scr}_{N+k-j+k-\ell} v_\ell \, v_k^{-1} \right )
            \\
           &= -\sum_{\ell=M}^{k-1} 
            \left ( -
 \sum_{j=M}^{k-1} v_k^{-1}\, v_j \, {q_\scr}_{N+k-j+k-\ell} \right ) \,v_\ell \, v_k^{-1} %\\ &
            = -\sum_{\ell=M}^{k-1}
              {q_\scr}_{N+k-j} \, v_\ell \, v_k^{-1}
           = {q_\scr}_N.
\end{align*}
\end{Proof}
Thus we may write $\ishinv_h v$ without ambiguity in the non-commutative case, \textit{i.e}
\begin{equation}
    \ishinv_h v = x^h \lquo v = x^h \rquo v.
\end{equation}

\subsection{Quotients from the Whole Shifted Inverse in $\Rofx$}
We consider computing the left and right quotients in $R[x]$ from the whole shifted inverse.
We have the following theorem.
\begin{Theorem}[Left and right quotients from the whole shifted inverse in $\Rofx$]
Let $u, v \in R[x]$, $R$ a ring, with $\pdegree v = k$ and $v_k$ invertible in $R$. Then for $h \ge \pdegree u$, 
\begin{align*}
    u\,\lquo v &=  \ishift_{-h} (\ishinv_h(v) \times u)\quad\text{and}
    \\
    u\,\rquo v &=  \ishift_{-h} (u \times \ishinv_h(v)) .
\end{align*}
\label{thm:r-poly-shinv-gives-quo}
\end{Theorem}
\begin{Proof}
Consider first the right quotient.  
It is sufficient to show
\[
u = \ishift_{-h} (u \times \ishinv_h v)  \times v + r_\scr
\]
for some $r_\scr$ with $\pdegree r_\scr < k$.  It is therefore sufficient to show
\begin{equation}
\ishift_{-k} u =  \ishift_{-k} \big (\ishift_{-h} (u \times \ishinv_h v)  \times v\big ).
\label{eq:shinv-rquo-to-show}
\end{equation}
We have
\begin{align}
(u \times \ishinv_h v) \times v 
            &= u \times ((x^h \rquo v) \times v) 
            \label{eq:shinv-rquo-mutatis-mutandis}
            \\
            &= u \times (x^h - \rho), \quad \rho = 0 \text{ or } \pdegree \rho < k
            \notag \\
            &= \ishift_h u - u \times \rho.
            \notag \\
\ishift_h u &= (u\times \ishinv_h v) \times v + u \times \rho.
            \label{eq:u-shifted}
\end{align}
Since $h \ge 0$, Theorem~\ref{thm:shift-cancel} applies and
equation \eqref{eq:u-shifted} gives
\begin{align*}
u &= \ishift_{-h} \big ((u \times \ishinv_h v) \times v \big) + \ishift_{-h}(u \times \rho)
\end{align*}
with the degree of $\ishift_{-h}(u \times \rho)$ less than  $k$. Therefore
\begin{align*}
\ishift_{-k} u &= \ishift_{-k -h} \big ((u \times \ishinv_h v) \times v \big)  \\
               &= \ishift_{-k} \big ( \ishift_{-h}(u \times \ishinv_h  v) \times v) \big ),
\end{align*}
by Theorem~\ref{thm:shift-factor}, and
we have shown equation~\eqref{eq:shinv-rquo-to-show} as required.
The proof for $\lquo$ replaces equation~\eqref{eq:shinv-rquo-mutatis-mutandis} with
\[
v \times (\ishinv_h v \times u) = (v \times (x^h \lquo v)) \times u
\]
and follows the same lines, \textit{mutatis mutandis}.
\end{Proof}
As in the commutative case,  it may be more efficient to compute only the top part of the product instead of computing the whole thing then shifting away part.
Now that we have shown that $\ishift$ and $\ishinv$ are well-defined for non-commutative $R[x]$, we next see that $\ishinv$ may be computed by our generic algorithm.

%%%%%%%%%%%%%%%%%%%%%%%%%%%%%%%%%%%%%%%%%%%%%%%%%%%%%%%%%%%%%%%%%%
\section{Generic Algorithm for the Whole Shifted Inverse}
\label{sec:generic-algo}
%%%%%%%%%%%%%%%%%%%%%%%%%%%%%%%%%%%%%%%%%%%%%%%%%%%%%%%%%%%%%%%%%%
Earlier work has shown how to compute $\ishinv$ efficiently for $\mathbb Z$, both for Euclidean domains $F[x]$, and generically~\cite{watt-issac-2023}.
The generic version shown here in Algorithm~\ref{algo:generic-whole-shifted-inverse}.  We justify below that it applies equally well to polynomials with non-commutative coefficients.
The algorithm operates on a ring $D$ that is
required to have a suitable $\ishift$ and certain other operations and properties must be defined.
For example, on $F[x]$, $F$ a field, these are
\begin{align*}
    \ishift_n u &= \begin{cases}
      u \cdot x^n & \text{if~} n \ge 0 \\
      u \iquo x^{-n} &\text{if~} n < 0
    \end{cases} \\
    \text{coeff}(u, i) &= u_i \\
    \Sc{Shinv0}(v) &= (1/v_k \,x - 1/v_k \cdot v_{k-1}\cdot 1/v_k,\; 2) \\
    \Sc{hasCarries} &= \text{false} \\
    \Sc{Mult}(a, b) &= ab \\
    \Sc{MultMod}(a,b,n) &= ab \irem x^n.
\end{align*}
The iterative step of Algorithm~\ref{algo:generic-whole-shifted-inverse} is given on line~\ref{line:step-body}.
Since $\Sc{D.PowDiff}$ computes $\ishift_h 1 - v \cdot w$, this line computes 
\begin{equation}
    \ishift_m w + \ishift_{2m-h}\big (w \cdot (\ishift_h 1 - v\cdot w)\big ).
    \label{eq:algo-step-expression}
\end{equation}
The shift operations are multiplications by powers of $x$, with $\ishift_h p = p x^h$. 
The the expressions involving $k, h, \ell$ and $m$ for shift amounts arise from multiplication by various powers of $x$ at different points in order to compute shorter polynomials when possible. Since $x$ commutes with all values, it is possible to accumulate these into single pre- and post- shifts.
With this in mind, the $R[x]$ operations $+$ and $\cdot$ ultimately compute the polynomial coefficients using the operations of $R$ and the order of the multiplicands in \eqref{eq:algo-step-expression} is exactly that of the 
Newton-Schulz iteration \eqref{eq:newton-schulz-iteration}.
The form of $\Sc{Shinv0}$ above is chosen so that it gives a suitable initial value for non-commutative polynomials.

The computational complexity of the $\Sc{Refine}$ methods of Algorithm~\ref{algo:generic-whole-shifted-inverse} may be summarized as follows:
The function $\Sc{D.Refine1}$ computes full-length values at each iteration
so has time complexity $O(\log(h-k) M(h))$ where $M(N)$ is the time complexity of multiplication.  The function $\Sc{D.Refine2}$ reduces the size of the values, computing only the necessary prefixes.
The function $\Sc{D.Refine3}$ reduces the size of some values further and achieves time complexity $O\big (\sum_{i=1}^{\log(h-k)} M(2^i)\big)$, which gives time complexity $O(M(N)), N = h-k$ for the purely theoretical $M(N) \in O(N \log N)$, for Sch\"{o}nhage-Strassen $M(N) \in O(N \log N \log \log N)$ and for $M(N) \in O(N^p), p > 0.$
\begin{algorithm}[t]
\caption{Generic $\Sc{Shinv}(v,h)$}
\label{algo:generic-whole-shifted-inverse}
\begin{algorithmic}[1]
\Require{$v \in D, h \in \mathbb Z_{>0}\;$where $0 < k = \iprec v - 1 < h$}
\Ensure{$\ishinv_{h} v \in D$}  
\Function{\Sc{D.Shinv\,}}{$v,h$\algosmallspace}
\LComment{Domain-specific initialization}
\State $(w, \ell) \gets \Sc{D.Shinv0}(v)$ \Comment{Initialize $w$ to $\ell$ correct places.}
\State \Return $\Sc{D.Refine}(v,h,k,w, \ell)$ \Comment{One of $\Sc{D.Refine1}, \Sc{D.Refine2}, \Sc{D.Refine3}$.}
\EndFunction
\LComment{Below, $g$ is the number of guard places and $d$ is the precision doubling shortfall.\algomedspace}
\Function{\Sc{D.Refine1\,}}{$v, h, k, w, \ell$\algosmallspace}
\State \algorithmicif\ $\Sc{D.HasCarries}$
    \algorithmicthen\ $g \gets 1; \; d \gets 1$
    \algorithmicelse\  $g \gets 0; \; d \gets 0$
\State $h \gets h + g$
\State $w \gets \mathrm D.\ishift_{h-k-\ell}(w)$ \Comment{Scale initial value to full length}
\While{$ h-k + 1 -d >  \ell$}
\label{line:refine1-stop-condition}
    \State $w \gets  \Sc{D.Step}(h, v, w, 0, \ell)$
    \State $\ell \gets   \min(2 \ell-d, h-k+1-d)$
    \Comment{Number of accurate digits}
\EndWhile
\State \Return $w$
\EndFunction
\Function{\Sc{D.Refine2\,}}{$v, h, k, w, \ell$\algomedspace}
\State \algorithmicif\ $\Sc{D.HasCarries}$
    \algorithmicthen\ $g \gets 2; \; d \gets 1$
    \algorithmicelse\  $g \gets 0; \; d \gets 0$
    \State $w \gets \mathrm D.\ishift_g w$
\While{$ h-k + 1 -d>  \ell$}
    \State $m \gets \min(h-k+1-\ell, \ell)$ \Comment{How much to grow}
    \State $w \gets  
      \mathrm D.\ishift_{-d} \; \Sc{D.Step}\big (k+\ell + m +d - 1 + g,\; v, \;w\;, m,\;\ell - g\big )$
    \State $\ell \gets \ell + m - d$
\EndWhile
\State \Return $w$
\EndFunction
\Function{\Sc{D.Refine3}\,}{$v, h, k, w, \ell$\algomedspace}
\State \algorithmicif\ $\Sc{D.HasCarries}$
    \algorithmicthen\ $g \gets 2; \; d \gets 1$
    \algorithmicelse\  $g \gets 0; \; d \gets 0$
\State $w \gets \Sc{D}.\ishift_g  w$
\While{$h-k+1-d > \ell$}
    \State $m \gets \min(h-k+1-\ell, \,\ell)$
    \State $s \gets \max(0, \; k - 2\ell + 1 - g)$
    \State $w \gets   \Sc{D}.\ishift_{-d} \big (
      \Sc{D.Step}\big (  k+\ell + m -s - 1 +d +g,\,
      \Sc{D}.\ishift_{-s} v,\, w, \, m,\, \ell -g\big )
      \big )$
    \State $\ell \gets \ell + m -d$  
\EndWhile
\State \Return $\Sc{D}.\ishift_{-g}(w)$
\EndFunction
\Function{\Sc{D.Step}\,}{$h, v, w, m, \ell$\algomedspace}
\State $ \Sc{D}.\ishift_m w \;\;+ 
   \Sc{D}.\ishift_{2m-h} \Sc{Mult}
       \big (
          w,
          \Sc{D.PowDiff}(v, w, h-m, \ell)
       \big )$
       \label{line:step-body}
\EndFunction
\LComment{Compute $\mathrm D.\ishift_h 1 - v w$ efficiently.\algomedspace}
\Function{\Sc{D.PowDiff}\,}{$v,w,h,\ell$}
\State   $c \gets \algorithmicif\ \Sc{D.HasCarries}$ \algorithmicthen\ 1 \algorithmicelse\ 0 % = peek in Maple code
\State $L \gets \Sc{D}.\iprec v + \Sc{D}.\iprec w \;- \ell + c$
\Comment{$c$ for coeff to peek}
\If {$v = 0 \vee w = 0 \vee L \ge h$}
   \State \Return $\Sc{D}.\ishift_h 1 - \Sc{D.Mult}(v,w)$
\Else   
   \State $P \gets \Sc{D.MultMod}(v, w, L)$
   \If {$\Sc{D.HasCarries} \wedge \Sc{D}.\text{coeff}(P, L-1) \ne 0$}
       \Return $\Sc{D}.\ishift_L 1 -  P$
   \Else\
       \Return $-P$
   \EndIf
\EndIf
\EndFunction
\end{algorithmic}
\end{algorithm}

%%%%%%%%%%%%%%%%%%%%%%%%%%%%%%%%%%%%%%%%%%%%%%%%%%%%%%%%%%%%%%%%%%
\section{Non-Commutative Polynomial Example}
\label{sec:r-poly-example}
%%%%%%%%%%%%%%%%%%%%%%%%%%%%%%%%%%%%%%%%%%%%%%%%%%%%%%%%%%%%%%%%%%
We give an example of computing left and right quotients via the whole shifted inverse with $R[x] = {F_{7}}^{2\times2}[x]$
using the algorithms of Sections~\ref{sec:r-poly-division} and~\ref{sec:generic-algo}.
Note that $R[x]$ is not a domain---there may be zero divisors, but it is easy enough to check for them.   
This example, and the one in Section~\ref{sec:r-skew-example}, were produced using the \texttt{Domains} package in Maple~\cite{monagan-domains}.  
The setup to use the \texttt{Domains} package for this example is
\begin{verbatim}
      with(Domains);
      F     := GaloisField(7);
      F2x2  := SquareMatrix(2, F);
      PF2x2 := DenseUnivariatePolynomial(F2x2, x);
\end{verbatim}
We start with
\begin{align*}
u &= \mtwo4661 x^5 + \mtwo2201 x^4 + \mtwo2113 x^3 + \mtwo2041 x^2
+ \mtwo3354 x + \mtwo4512,  \\
v &= \mtwo4345 x^2 + \mtwo5304 x + \mtwo1261.
\end{align*}
The whole 5-shifted inverse of $v$ is then
\begin{equation*}
\ishinv_5 v = \mtwo5434 x^3 + \mtwo6041 x^2 + \mtwo1022 x + \mtwo5163.
\end{equation*}
From this, the left and right quotients and remainders are computed to be
\begin{align*}
q_\scl &= \mtwo2611 x^3 + \mtwo6100 x^2 + \mtwo2033 x + \mtwo3100, &
r_\scl &= \mtwo1641 x + \mtwo1443, \\
q_\scr &= \mtwo3550 x^3 + \mtwo1115 x^2 + \mtwo0555 x + \mtwo4026, &
r_\scr &= \mtwo2021 x + \mtwo0456.
\end{align*}

Taking a larger example where $u$ has degree 100 and $v$ degree 10, 
$\Sc{D.Refine1}$ computes $\ishinv_{100} v$ with one guard digit in 6 steps with intermediate values of $w$ all of $\iprec$ 92. 
Methods $\Sc{D.Refine2}$ and $\Sc{D.Refine3}$ compute the same result also in 6 steps but with values of $w$ have $\iprec$ 4, 8, 16, 32, 64, 92 successively. Method $\Sc{D.Refine3}$ uses a shorter prefix of $v$ on the first iteration ($s = 3$).
The Maple code used for this example is given in Figure~\ref{fig:maple-generic-polynomial}.

%%%%%%%%%%%%%%%%%%%%%%%%%%%%%%%%%%%%%%%%%%%%%%%%%%%%%%%%%%%%%%%%%%
\spacetune{\pagebreak}
\section{Division in $R[x; \sigma, \delta]$}
\label{sec:r-skew-division}
%%%%%%%%%%%%%%%%%%%%%%%%%%%%%%%%%%%%%%%%%%%%%%%%%%%%%%%%%%%%%%%%%%
We now examine the more general case where the polynomial variable does not commute with coefficients.
For quotients and remainders to be defined, a notion of degree is required and we note that this leads immediately to Ore extensions, or skew polynomials.  After touching upon classical algorithms, we introduce the notions of left and right whole shifted inverse.  We note that the modified Newton-Schulz iteration 
may be used to compute whole shifted inverses, though in this case there is no benefit over classical division.  Finally, we show how left and right whole shifted inverses may be used to compute right and left quotients, each with only one multiplication.

\subsection{Definitions and Classical Algorithms}
Consider a ring of objects with elements from a ring $R$ extended by $x$, with $x$ not necessarily commuting with elements of $R$.   
By distributivity, any finite expression in this extended ring is equal to a sum of monomials, the monomials composed of products of elements of $R$ and $x$. 
To have a well-defined degree compatible with that of usual polynomials, it is required that
\begin{equation}
\forall \,r \in R \; \exists \,a, b, c, d \in R \;\text{ s.t. }\;
x r - r x = a x + b = x c + d.
\label{eq:skew-degree-condition}
\end{equation}
We call the elements of such a ring skew polynomials.
Condition~\eqref{eq:skew-degree-condition} implies that for all $r \in R$
there exist $\sigma(r), \delta(r) \in R$ such that 
\begin{equation}
x \, r = \sigma(r)\,  x + \delta(r).
\label{eq:skew-commutator}
\end{equation}
Therefore, to have well-defined notion of degree, the ring must be an Ore extension, $R[x; \delta, \sigma]$.  Ore studied these non-commutative polynomials almost a century ago~\cite{ore-1933} and overviews of Ore extensions in computer algebra are given in~\cite{abramov+-2005,bronstein+-1996}.  
The subject is viewed from a linear algebra perspective in~\cite{jacobson-1937} and the complexity of skew arithmetic is studied in~\cite{vdH-complexity-of-skew}.
The ring axioms of $R[x; \sigma, \delta]$ imply that $\sigma$ be an endomorphism on $R$ and $\delta$ be a $\sigma$-derivation, \textit{i.e.} for all $r, s \in R$
\begin{align*}
\delta(r+s)& =\delta(r)+\delta(s) &
\delta(r\cdot s) & = \sigma(r) \cdot \delta(s) + \delta(r)\cdot s.
\end{align*}
Different choices of $\sigma$ and $\delta$ allow skew polynomials to represent linear differential operators, linear difference operators, $q$-generalizations of these and other algebraic systems.

Condition~\eqref{eq:skew-commutator} implies that it is possible to write any skew polynomial as a sum of monomials with all the powers of $x$ on the right or all on the left.  We will use the notation $\rcoeff ui$ for coefficients of skew polynomials with all powers of the variable on the right and $\lcoeff ui$ for coefficients with all powers of the variable on the left, \textit{e.g.}
\begin{equation*}
    u = \sum_{i = 0}^h \rcoeff ui x^i = \sum_{i=0}^h x^i \lcoeff ui.
\end{equation*}

Algorithm~\ref{algo:classical-skew-polynomial-quotient} gives left and right classical division in $R[x; \sigma, \delta]$.  
As in Section~\ref{sec:r-poly-division}, $\times_\pi$ is multiplication with arguments permuted by $\pi$.   
When $\sigma(r) = r$,  $R[x;\sigma, \delta]$ is a differential ring,
usually denoted $R[x, \delta]$,
and
Algorithm~\ref{algo:classical-skew-polynomial-quotient} specializes to Algorithm~\ref{algo:classical-non-commutative-polynomial-quotient}.
The left division algorithm applies only when $\sigma$ is bijective.
If left division is of primary interest, 
start from \(r x = x \sigma^*(r) + \delta^*(r) \) 
instead of~\eqref{eq:skew-commutator} 
and work in the adjoint ring $R[x; \sigma^*, \delta^*]$.
\begin{algorithm}[t]
\caption{Classical division for $R[x;\sigma, \delta]$ with invertible $v_k$\algotinyspace}
\label{algo:classical-skew-polynomial-quotient}
\begin{algorithmic}[1]
\LComment{Compute $q$ and $r$ 
from $u$ of degree $h$  and 
$v$ of degree $k$
such that $u = q \times_\pi v + r.$\algomedspace\newline
The left division algorithm applies when $\sigma$ is bijective.}
\Function{\Sc{skewdiv}\,}{$u, v \in R[x; \sigma, \delta], \pi\in S_2, \Sc{qcoeff}$}\algomedspace
    \State $v^* \gets \mathrm{inv}~\rcoeff vk$
    \State $q \gets 0$; $r \gets u$
    \For{$i \gets h-k$ to $0$ by $-1$}
        \State $t \gets \Sc{qcoeff}(\rcoeff r{i+k}, v^*, i, k) \times x^i$
               \label{line:comp-t}
        \State $q \gets q + t$ ; $r \gets r - t \times_\pi v$
               \label{line:mult-t}
    \EndFor
    \State \Return (q, r)
\EndFunction
\LComment{Left division:~\, 
$(q_{\Sc l}, r_{\Sc l}) \gets \Sc{lskewdiv}(u,v) 
\Rightarrow  u = v \times q_{\Sc l} + r_{\Sc l}$
\algomedspace}
\State $\Sc{lskewdiv}(u, v) \mapsto 
    \Sc{skewdiv}\big ( u,\, v,\, (2\, 1),\;
        (a,b,n,k) \mapsto \sigma^{-k}(b \times a) \big )$
\LComment{Right division: 
$(q_{\Sc r}, r_{\Sc r}) \gets \Sc{rskewdiv}(u,v) 
\Rightarrow  u = q_{\Sc r} \times v + r_{\Sc r}$
\algomedspace}
\State $\Sc{rskewdiv}(u, v) \mapsto 
    \Sc{skewdiv}\big ( u,\, v,\, (1\, 2),\;
        (a,b,n,k) \mapsto a \times \sigma^n(b) \big )$
\end{algorithmic}
\end{algorithm}

Some care is needed in Algorithm~\ref{algo:classical-skew-polynomial-quotient} to avoid duplicating computation.
Notice that for $\Sc{rskewdiv}$ the application of $\Sc{qcoeff}$ on line~\ref{line:comp-t} requires $n$-fold application of $\sigma$ to $\iinv v_k$ and that the computation of $t\times_\pi v$ on line~\ref{line:mult-t} is $\text{coeff}(t)\, x^{i+k} \times v$.
The latter requires commuting $h-k$ powers of $x$ across $v$ over the course of the division.
Depending on the cost to compute $\sigma$, it may be useful to create an array of the values $\sigma^i(\iinv v_k)$ for $i$ from 0 to $h-k$.   
It is also possible to pre-compute and store the products $x^i \times v$, with $x^{i+1}\times v$ obtained from $x^i\times v$ by one application of~\eqref{eq:skew-commutator}. Then the $x^i\times v$ may be used in descending order in the \textbf{for} loop without re-computation.  Both of these pre-computations are performed in the Maple program for \verb+P[RDiv]+ shown in Figure~\ref{fig:maple_skew_poly}.

\subsection{Whole Shift and Inverse in $\Rskew$}
It is possible to define left and right analogs of the whole shift and whole shifted inverse for skew polynomials.   In general, the left and right operations give different values.

\begin{Definition}[Left and right whole $n$-shift in $\Rskew$]
\mbox{~}\newline
Given $u  = \in R[x; \sigma, \delta]$ and $n \in \mathbb Z$, 
the \emph{left} \emph{whole $n$-shift of $u$} is
\begin{equation*}
    \ilshift_{n,x} u = \sum_{i+n \ge 0} x^{i+n} \lcoeff ui,
\end{equation*}
the \emph{right} \emph{whole $n$-shift of $u$} is
\begin{equation*}
    \irshift_{n,x} u = \sum_{i+n \ge 0} \rcoeff ui x^{i+n}
\end{equation*}%
When $x$ is clear by context, we write $\ilshift_n u$ and $\irshift_n u$.
\end{Definition}

\begin{Definition}[Left and right whole $n$-shifted inverse in $\Rskew$]
\mbox{~}\newline
Given $n \in \NNInt$ and $v \in R[x; \sigma, \delta]$,
the \emph{left whole $n$-shifted inverse of $v$ with respect to $x$} is
\begin{equation*}
    \ilshinv_{n,x} v = x^n \lquo v
\end{equation*}
the \emph{right whole $n$-shifted inverse of $v$ with respect to $x$} is
\begin{equation*}
    \irshinv_{n,x} v = x^n \rquo v
\end{equation*}%
When $x$ is clear by context, we write $\ilshinv_n v$ and $\irshinv_n v$.
\end{Definition}

\subsubsection*{Modified Newton-Schulz Iteration}
For monic $v \in R[x; \sigma, \delta]$,  the whole shifted inverses may be computed using modified Newton-Schulz iterations with $g=1$ guard places as follows:
\begin{equation}
\begin{aligned}
{w_\scl}\iter 0 &= {w_\scr}\iter 0=  x^{h-k+g} - \rcoeff v{k-1} x^{h-k-1+g}\\
{w_\scl}\iter{i+1} &= {w_\scl}\iter i + \irshift_{-h} \big ( {w_\scl}\iter i \times (\irshift_h 1  - v \times {w_\scl}\iter i) \big), \\
{w_\scr}\iter{i+1} &= {w_\scr}\iter i + \,\ilshift_{-h} \big ( (\ilshift_h 1  - {w_\scr}\iter i \times v) \times {w_\scr}\iter i \big), \\
&\irshift_{-g} {w_\scl}\iter i \rightarrow \ilshinv_h v\\
&\ilshift_{-g} {w_\scr}\iter i \rightarrow \irshinv_h v.
\end{aligned}
\label{eq:skew-newton-schulz}
\end{equation}
These generalize $\Sc{D.Refine1}$ in Algorithm~\ref{algo:generic-whole-shifted-inverse}.
For $\Sc{D.Refine2}$ and $\Sc{D.Refine3}$, the shifts that reduce the size of intermediate expressions are combined into one pre- and one post-shift in $R[x]$.  But on $R[x;\sigma, \delta]$ we do not expect these simplifications of shift expressions to be legitimate.

Even though~\eqref{eq:skew-newton-schulz} \emph{can} be used to compute whole shifted inverses, it does not give any benefit over classical division.
In the special case of $R[x,\delta]$,
the multiplication by $v$ and then by $w$ make it so each iteration creates only one correct term,
so $h-k$ iterations are required rather than $\log_2(h-k)$.
In other skew polynomial rings, \textit{e.g.} linear difference operators, the iteration~\eqref{eq:skew-newton-schulz} can still converge, but with multiple iterations required for each degree of the quotient.   
It is therefore simpler to compute $\ilshinv$ and $\irshinv$ by classical division.

\subsection{Quotients from Whole Shifted Inverses in $\Rskew$}
It is possible to compute left and right quotients from the right and left whole shifted inverses in $R[x;\sigma,\delta]$.
Although computing whole shifted inverses is not asymptotically fast as it is in $R[x]$, once a whole shifted inverse is obtained it can
be used to compute multiple quotients and hence remainders, each requiring only one multiplication. 
This is useful, \textit{e.g.}, when working with differential ideals.   
In some cases this multiplication of skew polynomials is asymptotically fast~\cite{vanderhoeven-2002}.
\begin{Theorem}[Quotients from whole shifted inverses in $\Rskew$]~\newline
Let $u, v \in R[x; \sigma, \delta]$, with $R$ a ring, $k = \pdegree v$, $h = \pdegree u$, and $\rcoeff vk$ invertible in $R$.  Then 
\begin{align}
    u \rquo v &= \irshift_{-h} ( u \times \ilshinv_h v )
    \label{eq:skew-rquo} \\
    u \lquo v &= \ilshift_{-h} ( \irshinv_h v \times u).
    \label{eq:skew-lquo}
\end{align}
\label{thm:r-skew-quos-from-shinvs}
\end{Theorem}
\vspace{-\baselineskip}
\begin{Proof}
\newcommand{\uhat}{\ensuremath{\hat u}}
\newcommand{\qhat}{\ensuremath{\hat q}}
\newcommand{\rhat}{\ensuremath{\hat r}}
We first prove~\eqref{eq:skew-rquo}.  
For $h \ge k$, we proceed by induction on $h-k$.
Suppose $h-k = 0$.   
Since $ u - (\rcoeff uh \times 1/\rcoeff vk) \times v $ has no term of degree $h$, we have
\begin{equation*}
u \rquo v = \rcoeff uh \times 1/ \rcoeff vk.
\end{equation*}
On the other hand, when $h=k$, $\ilshinv_h v = 1/\rcoeff vk$ so 
\begin{equation*}
\irshift_{-h}(u\times \ilshinv_h v) = \rcoeff uh \times 1/\rcoeff vk
\end{equation*}
and \eqref{eq:skew-rquo} holds.
For the inductive step, we assume that~\eqref{eq:skew-rquo} holds for $h-k < N$.  
For $h-k=N$, let $u = q \times v + o(x^k)$ and let $Q$, $\qhat$ and $\uhat$ be given by
\begin{align*}
 u &= (Q x^{h-k} + \qhat) \times v + r, \quad\quad Q \in R,\;\; \qhat \in o(x^{h-k}),\;\; r \in o(x^k), \\
 \uhat &= u - Q x^{h-k} \times v.
\end{align*}
With this, $\uhat$ has degree at most $h-1$.  
The inductive hypothesis gives
\(
    \uhat \rquo v  = \irshift_{-h}(\uhat \times \ilshinv_h v).
\)
Therefore, 
\begin{align*}
\uhat = u - Q x^{h-k}\times v &= (\uhat \rquo v) \times v + \rhat,  \quad \rhat \in o(x^k) \\
   &= \irshift_{-h}(\uhat \times \ilshinv_h v) \times  v + \rhat \\
   \Rightarrow\quad
 u &= \big (\irshift_{-h}(\uhat \times \ilshinv_h v) + Q x^{h-k}\big) \times v + \rhat \\
   &= \irshift_{-h}(\uhat \times \ilshinv_h v + Q x^{2h-k}) \times v + \rhat.
\end{align*}
From this, we have
\begin{align*}
u \rquo v &= \irshift_{-h}(\uhat \times \ilshinv_h v + Q x^{2h-k}) \\
  &= \irshift_{-h}\big ( (u- Q x^{h-k} \times v) \times \ilshinv_h v + Q x^{2h-k}\big) \\
  &= \irshift_{-h}\big ( u\times \ilshinv_h v - Q x^{h-k} \times v \times \ilshinv_h v + Q x^{2h-k}\big) \\
  &= \irshift_{-h}\big ( u\times \ilshinv_h v - Q x^{h-k} \times v \times (x^h \lquo v) + Q x^{2h-k}\big) \\
  &= \irshift_{-h}\big ( u\times \ilshinv_h v - Q x^{h-k} \times (x^h + o(x^k)) + Q x^{2h-k}\big) \\
  &= \irshift_{-h}\big ( u\times \ilshinv_h v + Q \times o(x^h) \big)
  = \irshift_{-h}( u\times \ilshinv_h v).
\end{align*}
This completes the inductive step and the proof of~\eqref{eq:skew-rquo}.
% Since $\lcoeff vk = \sigma^k(\rcoeff vk)$ and $\sigma$ is an endomorphism, $\lcoeff vk$ is invertible 
Equation~\eqref{eq:skew-lquo} is proven as above, \textit{mutatis mutandis}.
\end{Proof}
As in the commutative case, it may be more efficient to compute only the required top part of the product in~\eqref{eq:skew-rquo} and~\eqref{eq:skew-lquo} rather than to compute the whole product and then shift by $-h$.

%%%%%%%%%%%%%%%%%%%%%%%%%%%%%%%%%%%%%%%%%%%%%%%%%%%%%%%%%%%%%%%%%%
\section{Skew Polynomial Examples}
\label{sec:r-skew-example}
\newcommand{\eqindent}{\hspace{2em}}
\newcommand{\Deriv}{\ensuremath{\,\partial}}
%%%%%%%%%%%%%%%%%%%%%%%%%%%%%%%%%%%%%%%%%%%%%%%%%%%%%%%%%%%%%%%%%%
\subsection{Differential Operators}
We take $F_{7}[y, \partial_y]$ as a first example of using whole shifted inverses to compute quotients of skew polynomials.  
We use Algorithm~\ref{algo:classical-skew-polynomial-quotient} to compute the left and right whole shifted inverses, and then Theorem~\ref{thm:r-skew-quos-from-shinvs} to obtain the quotients.
We start with  $u$ and $v$
\begin{align*}
u &= (3y+6) \partial_y^5 + (3y+1) \partial_y^4  + 6 y \partial_y^3 
+ 4 y \partial_y^2  + (2y+1) \partial_y  +  (2y+5)
\\
v &= 4 \partial_y^2 + (2y+ 5) \partial_y  + (4 y +  6).
\end{align*}
The whole shifted inverses $\ilshinv_5 v = \partial_y^5 \lquo v$ and $\irshinv_5 = \partial_y^5 \rquo v$ are computed by Algorithm~\ref{algo:classical-skew-polynomial-quotient}.
\begin{align*}
\ilshinv_5 &= 2 \partial_y^3 + (6y+1) \partial_y^2 
+ (4 y^2 + 4 y + 3) \partial_y  + (5 y^3 + y^2 + 3 y + 2)
\\
\irshinv_5 &=
  2 \partial_y^3
+ (6y+ 1) \partial_y^2 
+ (4 y^2 + 4 y + 5) \partial_y 
+ (5 y^3 + y^2 + y + 1).
\end{align*}
Then $q_\scl = \ilshift_{-5}(\irshinv_5 v \times u)$ and $q_\scr = \irshift_{-5}(u \times \ilshinv_5 v)$ so
\begin{align*}
q_\scl &=
  (6y+5) \partial_y^3
+ (4 y^2 + 3 y + 3) \partial_y^2 
+ (5 y^3 + 5 y^2 + 5) \partial_y \\&
+ (y^4 + 3 y^3 + 5 y^2 + 5 y + 2)
\\
r_\scl &=
  (5 y^5 + 4 y^4 + 3 y^3 + 6 y^2 + 4 y) \partial_y
+ (3 y^5 + 2 y^4 + y^3 + 5 y^2 + 5 )
\\[.8\baselineskip]
q_\scr &=
  (6 y+5) \partial_y^3
+ (4 y^2 + 3 y + 1) \partial_y^2 
+ (5 y^3 + 5 y^2 + 4 y + 3) \partial_y \\&
+ (y^4 + 3 y^3 + 5 y^2 + 3 y + 5 )
\\
r_\scr &=
  (5 y^5 + 4 y^4 + 6 y^3) \partial_y
+ (3 y^5 + 3 y^4 + 5 y^3 + y^2 + 4 y + 5 ).
\end{align*}
A proof-of-concept Maple implementation for generic skew polynomials is given in Figure~\ref{fig:maple_skew_poly}.
The program is to clarify any ambiguities without any serious attention to efficiency.
The setup for the above example is
\begin{verbatim}
    with(Domains):
    LinearOrdinaryDifferentialOperator :=
        (R, x) -> SkewPolynomial(R, x, r->r, R[Diff], r->r):

    F    := GaloisField(7):
    R    := DenseUnivariatePolynomial(F, 'y'): 
    Lodo := LinearOrdinaryDifferentialOperator(R, 'D[y]'):
\end{verbatim}
\spacetune{\pagebreak}
\subsection{Difference Operators}
We use linear ordinary difference operators as a second example, this time with $\sigma$ not being the identity.  We construct $F_7[y, \Delta_y]$ as $F_7[y][\Delta_y; E, E - 1]$.
As before, 
we use Algorithm~\ref{algo:classical-skew-polynomial-quotient} to compute the left and right whole shifted inverses, and then Theorem~\ref{thm:r-skew-quos-from-shinvs} to obtain the quotients.
We take $u$ and $v$ to be
\begin{align*}
u &= y \Delta_y^5 + (3 y + 6) \Delta_y^4  + (6 y + 5) \Delta_y^3 
+ 3 y \Delta_y^2  + (2 y + 1) \Delta_y  +  5 y 
\\
v &= 4 \Delta_y^2 + (6 y + 1) \Delta_y  + (6 y + 6).
\end{align*}
The whole shifted inverses $\ilshinv_5 v = \Delta_y^5 \lquo v$ and $\irshinv_5 = \Delta_y^5 \rquo v$ are computed by Algorithm~\ref{algo:classical-skew-polynomial-quotient}.
\begin{align*}
\ilshinv_5 &= 2 \Delta_y^3 + (4 y + 2) \Delta_y^2  
            + (y^2 + 4 y) \Delta_y  + (2 y^3 + 6 y^2 + y)
\\
\irshinv_5 &= 2 \Delta_y^3 + (4y + 1) \Delta_y^2 
            + (y^2 + 2) \Delta_y  + (2 y^3 + y^2 + 4 y + 1).
\end{align*}
Then $q_\scl = \ilshift_{-5}(\irshinv_5 v \times u)$ and $q_\scr = \irshift_{-5}(u \times \ilshinv_5 v)$ so
\begin{align*}
q_\scl &=
  (2 y + 3) \Delta_y^3
+ (4 y^2 + 3 y + 4) \Delta_y^2 
+ (y^3 + 5 y^2 + 6 y + 4) \Delta_y 
\\&
+ (2 y^4 + 6 y^3 + 4 y^2 + 4 y + 4)
\\
r_\scl &=
  (2 y^5 + 6 y^4 + 6 y^2 + 5 y + 3) \Delta_y
+ (2 y^5 + 2 y^4 + 4 y^3 + 2 y + 1)
\\
q_\scr &=
  2 y \Delta_y^3
+ (4 y^2 + 5) \Delta_y^2 
+ (y^3 + 5 y^2 + y + 6) \Delta_y 
+ (2 y^4 + 4 y^3 + 5 y + 1 )
\\
r_\scr &=
  (2 y^5 + 3 y^4 + 4 y^3 + y^2) \Delta_y
+ (2 y^5 + 6 y^4 + 5 y^3 + 3 y^2 + 5 y ).
\end{align*}
The Maple setup for this example is
\begin{verbatim}
    # Delta(f) acts as  subs(y=y+1, f) - f  for f in R
    LinearOrdinaryDifferenceOperator := proc(R, x, C)
        local E := R[ShiftOperator];
        SkewPolynomial(R, x, r->E(r,C[1]), r->R[`-`](E(r,C[1]),r),
                             r->E(r,C[`-`](C[1])));
    end:
      
    F    := GaloisField(7); 
    R    := DenseUnivariatePolynomial(F, 'y'); 
    Lodo := LinearOrdinaryDifferenceOperator(R, 'Delta[y]', F)
\end{verbatim}
\subsection{Difference Operators with Matrix Coefficients}
As a final example, we take quotients in $F_7^{2\times 2}[y, \Delta_y]$ to underscore the genericity of this method.  
\begin{align*}
u &= 
  \left ( \mtwo6011 y + \mtwo3020 \right ) \Delta_y^5
+ \left ( \mtwo4465 y + \mtwo3244 \right ) \Delta_y^4 
+ \left ( \mtwo4303 y + \mtwo1141 \right ) \Delta_y^3 
\\&
+ \left ( \mtwo0145 y + \mtwo3254 \right ) \Delta_y^2 
+ \left ( \mtwo0643 y + \mtwo0006 \right ) \Delta_y 
+ \left (  \mtwo5362 y  + \mtwo5212 \right )
\\
v &=\upstrut{5ex}
  \mtwo1526 \Delta_y^2
+ \left ( \mtwo1500 y + \mtwo4634 \right ) \Delta_y 
+ \left ( \mtwo2604 y  + \mtwo0312 \right )
\end{align*}
\spacetune{\setlength{\jot}{0.6\baselineskip}}
\begin{align*}
\ilshinv_5 &= 
  \mtwo2345 \Delta_y^3
+ \left ( \mtwo5030 y + \!\mtwo0412 \right ) \Delta_y^2 
+ \left ( \mtwo2040 y^2 + \!\mtwo3101 y +  \!\mtwo0244 \right ) \Delta_y 
\\&
+ \left ( \mtwo5030 y^3  + \mtwo4204 y^2 +  \mtwo2666 y +  \mtwo1266 \right )
\\[.3\baselineskip]
\irshinv_5 &= 
\mtwo2345 \Delta_y^3
+ \left ( \mtwo5030 y + \!\mtwo4422 \right ) \Delta_y^2 
+ \left ( \mtwo2040 y^2 + \!\mtwo2151 y +  \!\mtwo6002 \right ) \Delta_y 
\\&
+ \left ( \mtwo5030 y ^3  + \mtwo2234 y ^2  + \mtwo3554 y + \mtwo1331  \right )
\end{align*}
\begin{align*}
q_\scl &= 
  \left ( \mtwo1315 y + \mtwo3164 \right ) \Delta_y^3
+ \left ( \mtwo2040 y ^2 + + \mtwo4621 y + \mtwo2150 \right ) \Delta_y^2 
\\&
+ \left ( \mtwo5030 y ^3 + \mtwo4066 y ^2 + \mtwo2454 y + \mtwo0561 \right ) \Delta_y 
\\&
+ \left ( \mtwo2040 y^ 4  + \mtwo4326 y^ 3 + \mtwo1050 y^ 2 + \mtwo4315 y  + \mtwo5616 \right )
\\[.3\baselineskip]
r_\scl &= 
  \left ( \mtwo6000 y^5 + \mtwo6210 y^4 +  \mtwo6646 y^3 +  \mtwo2236 y^2 +  \mtwo2460 y +  \mtwo6520 \right ) \Delta_y
\\&
+ \left ( \mtwo0050 y^5  + \mtwo6034 y^4 +  \mtwo3236 y^3 +  \mtwo5130 y^2 +  \mtwo3646 y +  \mtwo2426 \right )
\end{align*}
\begin{align*}
q_\scr &= 
  \left ( \mtwo5461 y + \mtwo6246 \right ) \Delta_y^3
+ \left ( \mtwo2010 y^2 + \mtwo0060 y +  \mtwo5345 \right ) \Delta_y^2 
\\&
+ \left ( \mtwo5060 y^3 + \mtwo1602 y^2 +  \mtwo5514 y +  \mtwo5326 \right ) \Delta_y 
\\&
+ \left ( \mtwo2010 y^4  + \mtwo2556 y^3 +  \mtwo5243 y^2 +  \mtwo2211 y +  \mtwo2523 \right )
\\[.3\baselineskip]
r_\scr &= 
  \left ( \mtwo5462 y^5 + \mtwo1403 y^4 +  \mtwo4432 y^3 +  \mtwo1314 y^2 +  \mtwo3225 y +  \mtwo2645 \right ) \Delta_y
  \\&
+ \left ( \mtwo3251 y^5  + \mtwo3446 y^4 +  \mtwo3026 y^3 +  \mtwo6126 y^2 +  \mtwo3260 y +  \mtwo4013 \right )
\end{align*}
The Maple setup for this example is the same as for the previous example but with \verb+F := SquareMatrix(2, GaloisField(7))+.
%%%%%%%%%%%%%%%%%%%%%%%%%%%%%%%%%%%%%%%%%%%%%%%%%%%%%%%%%%%%%%%%%%
\section{Conclusions}
\label{sec:conclusions}
%%%%%%%%%%%%%%%%%%%%%%%%%%%%%%%%%%%%%%%%%%%%%%%%%%%%%%%%%%%%%%%%%%
We have extended earlier work on efficient computation of quotients in a generic setting to the case of non-commutative univariate polynomial rings.  
We have shown that
when the polynomial variable commutes with the coefficients,
the whole shift and whole shifted inverse are well-defined and they may be used to compute left and right quotients.  
The whole shifted inverse may be computed by a modified Newton method in exactly the same way as when the coefficients are commutative and the number of iterations is logarithmic in the degree of the result.
When the polynomial variable does not commute with the coefficients, left and right whole shifted inverses exist and may be computed by classical division.
Once a left or right whole shifted inverse is obtained, several right or left quotients with that divisor may be computed, each with a single multiplication.
%%%% Bibliography  %%%%%%%%%%%%%%%%%%%%%%%%%%%%%%%%%%%%%%%%%%%%%%%%%%%%%%%%%%
\bibliography{main}
\begin{figure}
\begin{scriptsize}
\begin{verbatim}
fshinv := proc (PR, method, h, v, perm)
    local R, x, k, vk, ivk, vkm1, w, ell, m, s, g, rmul, pmul, pshift, monom,
          step, refine, refine1, refine2, refine3;

    R      := PR[CoefficientRing];
    pmul   := (a, b)    -> PR[`*`](perm(a, b));
    rmul   := (a, b)    -> R [`*`](perm(a, b));
    monom  := (c, x, n) -> PR[`*`](PR[Polynom]([c]), PR[`^`](x, n));
    pshift := (n,v)     -> shift(PR, n, v);
    
    step   := proc(h, v, w, m, ell) 
        PR[`+`]( pshift(m,w), pshift(2*m-h,pmul(w,PR[`-`]( PR[`^`](x,h-m), pmul(v,w) ))) )
    end;

    refine1 := proc (v, h, k, w0, ell0) local m, s, w, ell;
        w := pshift(h-k-ell0+1, w0); ell := ell0;
        while ell < h-k+1 do
            w := step(h, v, w, 0, ell); ell := min(2*ell, h-k+1)
        od;
        w
    end;
    refine2 := proc (v, h, k, w0, ell0) local m, w, ell;
        w := w0; ell := ell0;
        while ell < h-k+1 do
            m := min(h-k+1-ell, ell);
            w := step(k+ell+m-1, v, w, m, ell); ell := ell+m
        od;
        w
    end;
    refine3 := proc (v, h, k, w0, ell0) local m, s, w, ell;
        w := w0; ell := ell0;
        while ell < h-k+1 do
            m := min(h-k+1-ell, ell); s := max(0, k-2*ell+1);
            w := step(k+ell+m-1-s, pshift(-s, v), w, m, ell); ell := ell+m
        od;
        w
    end;

    if   method = 1 then refine := refine1
    elif method = 2 then refine := refine2
    elif method = 3 then refine := refine3
    else error "Unknown method", method
    fi;
    
    x   := PR[Polynom]([R[0],R[1]]); k   := PR[Degree](v);
    vk  := PR[Lcoeff](v);            ivk := R[`^`](vk, -1);
    if   h < k then return 0
    elif k = 0 or h = k or v = monom(vk,x,k) then return monom(ivk,x,h-k)
    fi;
    vkm1 := PR[Coeff](v, k-1);
    w    := PR[Polynom]([rmul(ivk, rmul(R[`-`](vkm1), ivk)), ivk]);  ell := 2;
    g    := 1; # Assume all coeff rings need a guard digit
    pshift(-g, refine(v, h + g, k, w, ell))
end:

fdiv := proc (PR, method, u, v, perm) local mul, h, iv, q, r;
    mul := (a, b) -> PR[`*`](perm(a, b));
    h   := PR[Degree](u);
    iv  := fshinv(PR, method, h, v, perm);
    q   := shift(PR,-(h-k),mul(shift(PR,-k,u),iv)); # Need only top h-k terms
    r   := PR[`-`](u, mul(q, v));
    (q, r)
end:
lfdiv := (PR, method, u, v) -> fdiv(PR, method, u, v, (a,b)->(b,a)):
rfdiv := (PR, method, u, v) -> fdiv(PR, method, u, v, (a,b)->(a,b)):
\end{verbatim}
\end{scriptsize}
\vspace{-0.2\baselineskip}
\caption{Maple code for fast generic polynomial $\ishinv$ and left and right division}
\label{fig:maple-generic-polynomial}
\end{figure}
\begin{figure}[t]
\begin{scriptsize}
\begin{verbatim}
SkewPolynomial := proc (R, x, sigma, delta, sigmaInv)
    local P, deltaStar, mult2, MultVarOnLeft, MultVarOnRight;

    # Table to contain the operations.
    P := DenseUnivariatePolynomial(R, x);

    # If x*r = sigma(r)*x + delta(r), then
    #    r*x = x*sigmaInv(r) - delta(sigmaInv(r)) = x*sigmaInv(r) + deltaStar(r)
    deltaStar := r -> R[`-`](delta(sigmaInv(r)));

    P[DomainName]:= 'SkewPolynomial';
    P[Categories]:= P[Categories] minus {CommutativeRing,IntegralDomain};
    P[Properties]:= P[Properties] minus {Commutative(`*`)};

    P[ThetaOp]  := P[Polynom]([R[0], R[1]]);   # The variable as skew polynomial.

    P[Apply] := proc(ell, p) local i, pi, result;   # Apply a skew polynomial as an operator.
        pi     := p;   # delta^i (p)
        result := R[`*`](P[Coeff](ell, 0), pi);
        for i to P[Degree](ell) do  # For Maple, for loop default from is 1.
            pi     := delta(pi);
            result := R[`+`](result, R[`*`](P[Coeff](ell, i), pi))
        od;
        result
    end:

    P[`^`] := proc(a0, n0) local a, n, p;   # Binary powering
        a := a0; n := n0; p := P[1];
        while n > 0 do
          if irem(n,2) = 1 then p := P[`*`](p, a) fi; a := P[`*`](a, a); n := iquo(n,2);
        od;
        p
    end:

    P[`*`] := proc() local i, p;   # N-ary product
        p := P[1]; for i to nargs do p := mult2(p, args[i]) od; p
    end:
    mult2 := proc(a, b) local s, i, ai, xib;   # Binary product
        xib := b;  ai := P[Coeff](a,0);
        s   := P[Map](c->R[`*`](ai, c), xib);
        for i to P[Degree](a) do
            xib := MultVarOnLeft(xib); ai := P[Coeff](a, i);
            s   := P[`+`](s, P[Map](c->R[`*`](ai,c), xib));
        od;
        s
    end:
    
    # Compute x*b as polynomial with powers on right.
    # x*sum(b[i]*x^i, i=0..degb) = sum(sigma(b[i])*x^(i+1) + delta(b[i])*x^i, i=0..degb)
    MultVarOnLeft := proc(b) local cl, slist, dlist;
        cl    := P[ListCoeffs](b);
        slist := [ R[0], op(map(sigma, cl)) ]; dlist := [ op(map(delta, cl)), R[0] ];
        P[Polynom](zip(R[`+`], slist, dlist));
    end:
    # Compute b*x as polynomial with powers on left.
    # sum(x^i*b[i], i=0..degb)*x = sum(x^(i+1)*sigmaInv(b[i]) + deltaStar(b[i])*x^i, i=0..degb)
    MultVarOnRight := proc(b) local cl, slist, dlist;
        cl    := P[ListCoeffs](b);
        slist := [ R[0], op(map(sigmaInv, cl)) ]; dlist := [ op(map(deltaStar, cl)), R[0] ];
        P[Polynom](zip(R[`+`], slist, dlist));
    end:
 
    # Continued in Part 2...
\end{verbatim}
\end{scriptsize}
\caption{Maple code for generic skew polynomials (Part 1)}
\end{figure}
\begin{figure}[t]\ContinuedFloat
\begin{scriptsize}
\begin{verbatim}
    # ... continued from Part 1.

    # For v = sum(vr_i x^i, i = 0..k) = sum(x^i vl_i, i = 0..k)
    # return polynomial with vl_i, interpreting powers as on left,
    # abusing the representation of output.
    P[ConvertToAdjointForm] := proc(v) local v_adj, i, rci, rcip;
        v_adj := P[0];
        for i from P[Degree](v) to 0 by -1 do
            rci   := P[Polynom]([P[Coeff](v,i)]);
            v_adj := P[`+`](v_adj, (MultVarOnRight@@i)(rci));
        od;
        v_adj
    end:

    # For v = sum(x^i vl_i, i = 0..k) = sum(x^i vr_i, i = 0..k)
    # return polynomial with vr_i, interpreting powers as on right,
    # abusing the representation of input.
    P[ConvertFromAdjointForm] := proc(v_adj) local v, i, rci;
        v := P[0];  
        for i from 0 to P[Degree](v_adj) do
            rci := P[Polynom]([P[Coeff](v_adj,i)]);
            v   := P[`+`](v, (MultVarOnLeft@@i)(rci))
        od;
        v
    end:

    # Shift by power on left.
    P[LShift] := proc(n, v0) local v, shv, i, k;
        v := P[ConvertToAdjointForm](v0); k := P[Degree](v);
        if k + n < 0 then shv := P[0]
        elif n < 0 then shv := P[Polynom]([seq(P[Coeff](v,i), i = -n..k)])
        else shv := P[Polynom]([seq(R[0], i=1..n), seq(P[Coeff](v,i), i=0..k)])
        fi;
        P[ConvertFromAdjointForm](shv)
    end:

    # Shift by power on right.
    P[RShift] := proc(n, v) local i, k;
        k  := P[Degree](v);
        if k + n < 0 then P[0]
        elif n < 0 then P[Polynom]([seq(P[Coeff](v,i), i = -n..k)])
        else P[Polynom]([seq(R[0], i=1..n), seq(P[Coeff](v,i), i=0..k)])
        fi
    end:

    # Quotient and remainder
    P[GDiv] := proc(perm, qfun) proc (u, v) local h, k, x, ivk, t, q, r, i, qi; 
        x   := P[Polynom]([R[0], R[1]]); ivk := R[Inv](P[Lcoeff](v));
        h   := P[Degree](u); k := P[Degree](v);
        q   := P[0];         r := u;
        for i from h - k by -1 to 0 do
            qi := qfun(P[Coeff](r,i+k), ivk, i, k);
            t  := P[`*`](P[Constant](qi), P[`^`](x,i));
            q  := P[`+`](q, t);
            r  := P[`-`](r, P[`*`](perm(t, v)));
        od;
        (q, r)
    end end:
    P[RDiv0] := P[GDiv](rperm, (u,iv,n,k)->R[`*`](u,(sigma@@n)(iv)));
    P[LDiv]  := P[GDiv](lperm, (u,iv,n,k)->(sigmaInv@@k)(R[`*`](iv,u)));

    # Continued in Part 3...
\end{verbatim}
\end{scriptsize}
\caption{Maple code for generic skew polynomials (Part 2)}
\end{figure}
\begin{figure}[t]\ContinuedFloat
\begin{scriptsize}
\begin{verbatim}
    # ... continued from Part 2.
    
    # A slightly less repetitive RDiv.
    P[RDiv] := proc (u, v) local h, k, x, ivk, sigma_ivk_i, x_i_v, q, r, i, qi;
        x   := P[Polynom]([R[0], R[1]]); ivk := R[Inv](P[Lcoeff](v));
        h   := P[Degree](u); k := P[Degree](v);

        # Precompute sigma^i(ivk) and x^i*v for required i.
        sigma_ivk_i[0] := ivk;
        for i from 1 to h-k do sigma_ivk_i[i] := sigma(sigma_ivk_i[i-1]); od;
        x_i_v[0] := v;
        for i from 1 to h-k do x_i_v[i]       := P[`*`](x, x_i_v[i-1]) od;

        q   := P[0];   r := u;
        for i from h - k by -1 to 0 do
            qi := P[Constant](R[`*`](P[Coeff](r, i+k), sigma_ivk_i[i]));
            q  := P[`+`](q, P[`*`](qi, P[`^`](x,i)));
            r  := P[`-`](r, P[`*`](qi, x_i_v[i]));
        od;
        (q, r)
    end:

    # Needed for some versions of Maple.
    P[0] := P[Polynom]([R[0]]);
    P[1] := P[Polynom]([R[1]]);  
    P[`-`] := proc()
        local nb := P[Polynom](map(c-> R[`-`](c), P[ListCoeffs](args[nargs])));
        if nargs = 1 then nb else P[`+`](args[1], nb) fi
    end:

    # Return the table
    P
end:
\end{verbatim}
\end{scriptsize}
\caption{Maple code for generic skew polynomials (Part 3)}
\label{fig:maple_skew_poly}
\vspace{8cm}
\end{figure}
\end{document}